# The Joint COntrols Project Framework

M. Gonzalez-Berges on behalf of the Framework Team
*CERN, CH-1211, Geneve 23, Switzerland*

The Framework is one of the subprojects of the Joint COntrols Project (JCOP), which is collaboration between the four LHC experiments and CERN. By sharing development, this will reduce the overall effort required to build and maintain the experiment control systems. As such, the main aim of the Framework is to deliver a common set of software components, tools and guidelines that can be used by the four LHC experiments to build their control systems. Although commercial components are used wherever possible, further added value is obtained by customisation for HEP-specific applications. The supervisory layer of the Framework is based on the SCADA tool PVSS, which was selected after a detailed evaluation. This is integrated with the front-end layer via both OPC (OLE for Process Control), an industrial standard, and the CERN-developed DIM (Distributed Information Management System) protocol. Several components are already in production and being used by running fixed-target experiments at CERN as well as for the LHC experiment test beams. The paper will give an overview of the key concepts behind the project as well as the state of the current development and future plans.

## 1. INTRODUCTION

Given the scale, as well as the disperse location of teams working for the LHC experiments, there is enormous scope for duplication of work. Furthermore, the integration of the various pieces built separately is also a major issue. To highlight this, the four LHC experiments will have in total about 50 different sub-detectors many of which are comprised of multiple teams working in parallel. However as far as controls are concerned, these various groups will in many cases be using similar equipment and require very similar functionality.

The Joint COntrols Project [1] was set up in the end of 1997 to address these issues. As such it is intended to address the common aspects of the LHC experiments' Detector Control System (DCS) to reduce duplication and to ease integration. JCOP is organized in subprojects, of which the main ones are the Framework, the Gas Control Systems and the Detector Safety System (DSS).

The Framework is meant to provide a set of components to ease the development. The first design was proposed by the Architecture Working Group in 2001 [3]. After that the requirements and design have been extended and refined within the Framework Working Group, the body that drives the development process.

The project main aims are to reduce the development effort, by reusing common components and hiding the complexity of the underlying tools, and also to obtain a homogeneous control system that will ease the operation and maintenance of the experiments during their life span.

The approach that has been followed is to provide a high level of abstraction giving an interface for non-experts. As such, the required knowledge of the technologies involved is reduced. Industrial components are customized and extended to cope with the requirements of our domain. A set of components is available from which users can select and then mix and match to build their application. Since only components of general interest to the experiments are covered by the project, there is an easy mechanism to extend the Framework to meet experiment specific need. This is achieved by following well-defined interfaces.

## 2. ARCHITECTURE

A typical control system for an experiment is organized in two main layers: the front-end layer and the supervision layer. The front-end layer is responsible for giving access to the equipment, whilst the supervision offers an interface to the different human users and high-level operations of the experiment. The JCOP Framework is focused on providing components to build the supervision layer.

Figure 1 shows where the Framework sits within the architecture of a typical control system.

The supervision layer is based on the commercial SCADA tool PVSS [4], chosen after a long evaluation process. This is complemented with other packages (e.g. database access, Finite State Machines). The Framework customizes these general tools providing components of interest for the HEP environment. An overview of these components is given in the next section. Both Windows (2000, XP) and Linux (Red Hat) platforms are supported.

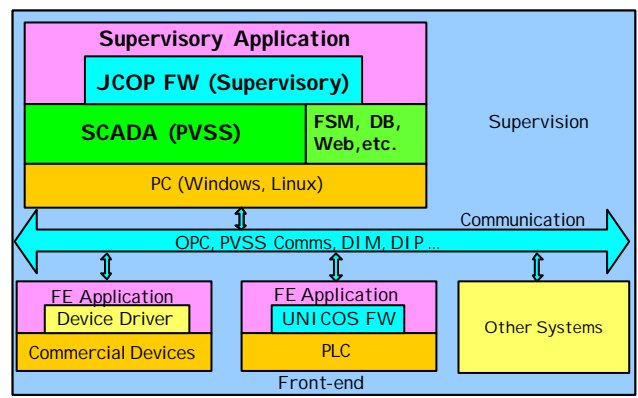

Figure 1. Typical architecture of a control system.

The DIM (Distributed Information Management System) and the OPC (OLE for Process Control) are used as the main protocols for the middleware that interfaces the supervision and front-end layers. The PVSS internal protocol can be used if there is a need to develop a specific driver. For the exchange of information with other CERN systems like the LHC machine or the Technical Services, the Data Interchange Protocol (DIP), currently being defined, will be used.

Although there are not many developments within the front-end layer, the Framework makes sure that everything down to the real equipment can be integrated coherently. When PLCs are used, the front-end part of the UNICOS[1] Framework [5] is recommended.

## 3. FRAMEWORK COMPONENTS

The different Framework components are described in the following sections.

### 3.1.1. Guidelines

The first deliverable that is provided by the Framework is a set of guidelines and conventions meant to be followed by all the developers either of the Framework or of the final control system. The guidelines contain chapters for: look & feel to use when designing panels, alarm classes that should be used and the meaning of each of them, coding rules, naming conventions, system integration, configuration management, exception handling, organization of files in the project, etc.

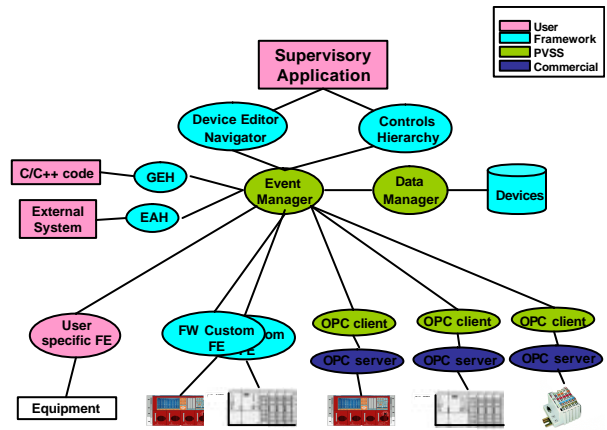

Figure 2. Framework Software Components.

### 3.1.2. Devices

A device can represent a piece of equipment (e.g. a high voltage channel) or a logical entity (e.g. a group of channels). In the supervision layer, a device consists of one or more data point types, a library that encapsulates the knowledge to manage the device instances, and a set of panels to provide an interface for the user. The front-end layer is specific for each application. The only constraint at this layer is to have an interface to one of the protocols supported within the Framework. Preference is given to the use of front-end software coming directly from the manufacturer of the hardware that is controlled (e.g. OPC server for CAEN equipment).

The following devices are currently at the disposal of the users:

- Generic Analog-Digital Devices

This provides basic functionality to deal with inputs and outputs that can reside in a PLC, a fieldbus module, an I/O card, etc.

- CAEN Power Supplies

The models that are currently in use at CERN are supported (SY127, SY403, SY527 and SY1527)

- Wiener Power Supplies

These crates have a CAN bus interface. All the models that support the standard Wiener protocol are included.

- Wiener Fan Tray

They follow a similar interface to the Wiener PS.

- PS and SPS[2] machine data server

Data from the CERN accelerators and their beamlines is made available for usage in the testbeam activities.

- Embedded Local Monitor Board

The ELMB [8] is a CERN developed general purpose input/output device. It is radiation tolerant and comes with a CAN bus interface.

- Logical Node

This device is a container for other device instances. Once grouped together, the set of instances offers the same interface as a single instance.

When the above list is not enough one can extend the Framework capabilities with the inclusion of a new device type. There exists an interface that has to be followed for this purpose. In addition, a template with common required functionality is provided to ease the process.

### 3.1.3. Tools

The following tools are currently at the disposal of the users:

- Device Editor and Navigator

This is the main interface to the Framework. One can configure and operate at a low level all the devices of the DCS. Simple device hierarchies can be built to give a logical structure to the DCS.

---

[1] UNified Industrial COntrol System. Controls Framework developed at CERN originally intended for the Cryogenics of the LHC accelerator and the experiments. Its usage has been extended to other project like the Gas Control Systems for the LHC experiments. For the supervision layer, PVSS is used, and the front end and control layers are based on PLCs.

[2] Proton Synchrotron and Super Proton Synchrotron.

System management actions like installation or login can be performed.

- Controls Hierarchy

Provides a high level view of the experiment. Includes modeling of the device behaviour by building a hierarchy of Finite State Machines (FSMs). As a general rule, only FSMs can take automatic actions in the system. The actions at a given level depend on the reported status of the layers below. More details can be found in [6] and [7].

- External Alarm Handler (EAH)

Alarms that have generated in a non-PVSS system can be injected into PVSS with the External Alarm Handler.

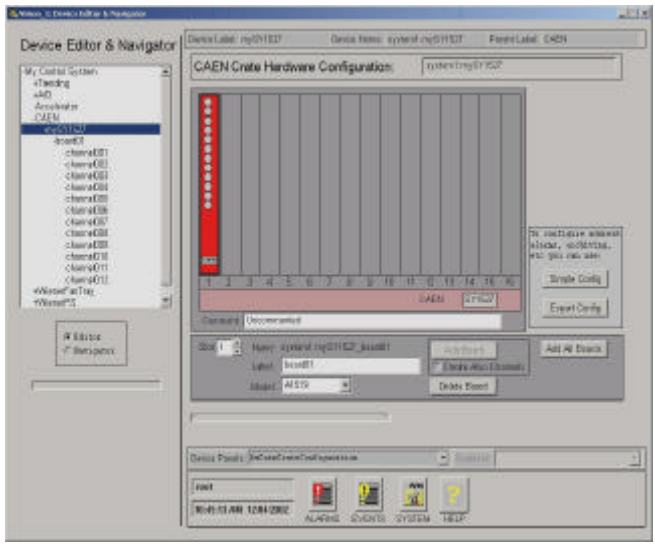

Figure 3. The Device Editor and Navigator.

- Trending Tool

PVSS provides a mechanism to plot process variables on a trend widget. The current data as well as the history of the variable can be displayed. The Trending Tool extends this features by providing a simple Framework oriented configuration of the displays, the inclusion of pages of plots, templates that are instantiated at run time with a given set of parameters, and the possibility to build trees of pages and plots. An example page is shown in Figure 4.

- Generic External Handler (GEH)

When performance is an issue or when already existing code has to be integrated, one can use the Generic External Handler to incorporate C/C++ code to PVSS panels and scripts.

- Mass Configuration.

The current release includes functionality to create and delete groups of devices selected according to different criteria (e.g. name, type).

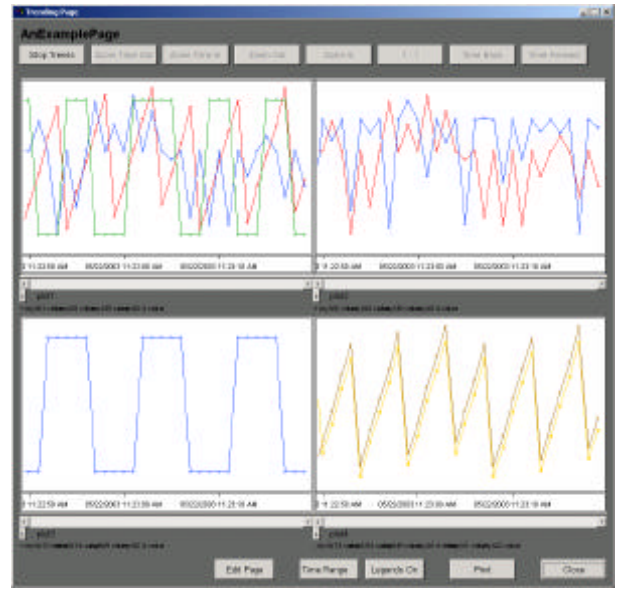

Figure 4. A Trending Tool page.

- Component Installer

As already mentioned, the Framework is developed and distributed in a modular way. The user can choose which components to install. Managing the versions of a given component is also an important issue.

### 3.1.4. Other

A set of libraries to deal with common operations (e.g. list manipulation), a catalogue of symbols that can be reused when building graphical user interfaces, and examples for commonly needed features are also available.

Recently, a tutorial has been written to help the users to familiarize themselves with the use of the Framework. This is distributed as a handout and can also be followed as a one-day course as part of CERN's Training Service.

## 4. THE DEVELOPMENT PROCESS

The development of any Framework component is driven by the Framework Working Group. Representatives from the four LHC experiments participate in the discussions. The group meets regularly every two weeks to identify new components and set priorities. Once a new development has been agreed, a short requirements document is written and reviewed. After that a design is proposed that it also reviewed by the FW WG. The implementation, testing and documentation follow later. The whole process is iterative to avoid diverging from the users' needs.

All of the material produced, either documentation or code, is kept under configuration management in a CVS repository. All of the deliverables are made available to users through the project web pages.

## 5. CURRENT DEVELOPMENTS

Apart from the extension with new features of the components that are already in production, at the moment, our efforts are centered on linking the DCS with databases. On one side, a configuration database is needed to keep different working versions of the system. On the other side, data coming from the experiment have to be stored in the Conditions Database for further analysis.

The Configuration Tool will deal with three types of information: system information (e.g. which processes run on each machine, versions of packages), static data (e.g. device instances, addresses) and device settings, also called recipes (e.g. values of parameters, alarm limits, archiving parameters). All of it will be stored in an external database. The experiments will be using a variety of databases, so a database independent mechanism is required. After an evaluation of the XML technologies (DTD, XML schema, XSLT), we have decided to go for ODBC, because it is a more established solution. Access from Linux is available by means of the Qt library [10].

Data gathered by the DCS and needed for later use can be stored in local archive files, or sent directly to the Conditions Database. For debugging of the DCS the data can be stored in PVSS archive files, and afterwards moved to CASTOR if long term storage is required. For offline analysis data has to be stored in the Conditions Database for which there is an agreement to have a common interface [9].

A way to introduce Access Control is being studied. There is also work going on to include ISEG Power Supplies as new devices. The equipment will be accessed by means of an OPC server provided by the company.

## 6. USERS

Interestingly, although developed for the LHC experiments, the Framework was first used by the fixed-target experiments (COMPASS, HARP and NA60). There has been a mutual benefit, on the one side they got a package that helped them to meet their short timescales for development, on the other we got help for the debugging and proof of concept of some of the Framework components.

Our main users are the LHC experiments and testbeams. We can mention as examples the following projects where the Framework is in use: ALICE/HMPID, TPC; ATLAS/ELMB, Muon; CMS/Muon DT; LHCb/Central testbeam.

The Gamma Irradiation Facility (GIF) is another of our users.

We keep a close collaboration with the UNICOS Framework [5] team to share the development of common components.

The LHC Gas Systems are currently using both, the JCOP and UNICOS Frameworks.

It is important to remark that in some of the above cases people with no previous experience in control systems have been able to develop a complete DCS after a short initial learning phase.

## 7. CONCLUSIONS

Although the JCOP Framework is still in its development phase, it is a mature product that is being used widely at CERN. The current supported devices and tools cover most of the common needs for the LHC experiments. Several other components are planned for the next years, with emphasis on databases for configuration and conditions during this year.

## Acknowledgements

The author would like to thank their many staff colleagues, students and visitors to CERN who have contributed in no small way to the various developments.